\documentclass[journal,comsoc]{IEEEtran}

\usepackage[colorlinks]{hyperref} 
\usepackage[T1]{fontenc}% optional T1 font encoding

\usepackage{graphicx} 							% 	inserting images
\usepackage{hyperref} 							% 	PDF links
\usepackage{mathrsfs}
\usepackage[noadjust]{cite}
\usepackage{amsmath}

\begin{document}

\title{Combining and Steganography of 3D Face Textures}

\author{Mohsen Moradi and Mohammad-Reza Rafsanjani-Sadeghi
\thanks{M. Moradi and M.-R. Sadeghi are with the Department of Mathematics, Amirkabir University of Technology - Tehran Polytechnic , Tehran, Iran, ( e-mail:
m.moradi8111/msadeghi@aut.ac.ir ).
}}

\maketitle

\begin{abstract}
One of the serious issues in communication between people is hiding information from others, and the best way for this is deceiving them. Since nowadays face images are mostly used in a three-dimensional format, in this paper we are going to steganography 3-D face images, detecting which by curious people will be impossible. As in detecting the face, only its texture is important, we separate texture from shape matrices, for eliminating half of the extra information, steganography is done only for face texture, and for reconstructing the 3-D face, we can use any other shape. Moreover, we will indicate that, by using two textures, how two 3-D faces can be combined. For a complete description of the process, first, 2-D faces are used as an input for building 3-D faces, and then 3-D face and texture matrices are extracted separately from the constructed 3-D face. Finally, 3-D textures are hidden within other images.
\end{abstract}


\IEEEpeerreviewmaketitle

\section{Introduction}

\IEEEPARstart{I}{n} cryptography, encrypted message is at the center of people's attention, and its security is based on the difficulties on access to the message key. In steganography, other's unawareness is used for hiding the message to send it in the safest way. For this reason, first, the essential information of one image which we want to send to others will be embedded in another host image in a way that others cannot understand outward discrepancy of the initial host image and the embedded host image, therefore legal receiver can extract and reconstruct the initial embedded image. The most important researches conducted in this area are \cite{Blanz1999, Zhu, Bas, Kittler, Patel}.\\

Embedding capacity and visual quality are the two essential parameters in the stego or cover images \cite{Chang}. Embedding capacity refers maximum amount of secretive message which can be embedded in the host image, and visual quality is embedding a message in the host image in a way that human eye cannot notice any difference between the new form and the original one. One criterion which is usually used for evaluating visual quality is the peak of the signal to noise ratio (PSNR) between stego image and original host image and expressed in dB unit. The bigger the PSNR, the higher visual quality of stego image. In other words, it is more difficult for the eye to detect stego image than to do so for host image \cite{Johnson}.\\.

In this paper, to construct the 3-D image of the intended face for steganography within other images first, a colored 2-D face image will be processed using Basel database as \cite{Paysan} to the 3-D model is fitted. 3-D face and texture matrices are extracted separately from the constructed 3-D face. Now, if this 3-D texture is used for any other 3-D shapes, it is possible to recognize the person as well. For this reason, in steganography, the shape matrix is not too important and we only use texture matrix. A wavelet-based watermarking algorithm is used to enhance the secrecy. By using singular value decomposition (SVD) and discrete wavelet transform (DWT), the information of a 3-D texture of an image will be hidden in a label image. This labeled image can be any 2-D image e.g., a fingerprint, a shape, or a texture of other 3-D images. First, label image is converted to the frequency domain, and SVD is used on both of the original cover image and 3-D texture. In the following, the two obtained singular values will be replaced with each other.

\section{	Morphable model}

At first, 3-D morphable models had been introduced by Blanz and Vetter \cite{Blanz1999, Blanz2003}. They were applied successfully in computer's images and graphics. A 3-DMM includes separate shape and texture models which by themselves can build each person's shape and texture's changes, respectively. An 3-DMM is used for one group of 3-D scanned images.\\

Constructing 3-D models is very hard and time-consuming which Paysan and his colleagues provided their 3-DMM for public using; Basel model uses 3-D models of 100 men and 100 women who ranged from 8 to 64 \cite{Paysan}.\\
An Iterative Multiresolution Dense 3-D Registration which has been introduced by Rodriguez \cite{Rodriguez} will be studied in this section. Suppose that the ith vertex of registered image be in the $(x_{i}, y_{i}, z_{i})$ point and its RGB color is $(R_{i}, G_{i}, B_{i})$. It will be assumed that for any face N 3-D points had been registered; therefore, one registered face in the face and texture language can be shown as:

\begin{equation}\label{equation1}
S^{'}=
  \begin{pmatrix}
    x_{1} , \cdots , x_{N} \\
    y_{1} , \cdots , y_{N} \\
    z_{1} , \cdots , z_{N} 
  \end{pmatrix}
\end{equation}

\begin{equation}\label{equation2}
T^{'}=
  \begin{pmatrix}
    R_{1} , \cdots , R_{N} \\
    G_{1} , \cdots , G_{N} \\
    B_{1} , \cdots , B_{N} 

  \end{pmatrix}.
\end{equation}

Also, these points can be shown by one row or column vector of 3N length. For example, the vertical vector of shape and texture are:
\begin{equation}
S^{'}=
  \begin{pmatrix}
    x_{1} , \cdots , x_{N},     y_{1} , \cdots , y_{N} ,    z_{1} , \cdots , z_{N} 

  \end{pmatrix}^{T},
\end{equation}
\begin{equation}
T^{'}=
  \begin{pmatrix}
    R_{1} , \cdots , R_{N} ,    G_{1} , \cdots , G_{N} ,    B_{1} , \cdots , B_{N} 

  \end{pmatrix}^{T}
\end{equation}
which N is the number of registered faces. We suppose that these two kinds of point's display are equivalent, and we will use any of these two formats when needed.  \\

Now, for a database which has 200 shapes and texture matrices, a linear combination for shapes and textures will be used. These linear combinations can be as follows:

\begin{equation} \label{equation5}
S= \sum_{i=1}^{200} \alpha _{i}. S_{i}  \ \ \ \ \ \ \ \ \ \ \   T= \sum_{i=1}^{200} \beta _{i} .T_{i} \ .
\end{equation}

It is almost impossible that these combinations, despite the fact that they include all the possible faces, be similar to a real face. If a convex combination ($\sum \alpha_{i}=1, \sum \beta _{i}=1 , \alpha_{i}, \beta _{i} \in [0,1] $) of faces be supposed, a face can also be obtained, but again, it is not possible that the points which are far from convex area form actual face points. Therefore, for any vector, a coefficient is needed to be allocated to a probability distribution for the description of a face. This probability is modeled by a Gaussian distribution, in which shapes and textures are decorrelated, with a diagonal matrix. Suppose a Gaussian distribution allows that subspace of face to be estimated by a smaller set of orthogonal basis vectors which is calculated using principle component analysis (PCA) of the test samples.  \\

Principle component analysis is a statistical tool which transform shape or texture so that covariance matrix will be diagonal (it means data are decorrelated). In this section, using PCA for the shape will be studied. Using it for the texture is done in a similar way. PCA is a transform in the vector space which is used mostly for decreasing the dimension of the data sets. At first, principle component analysis had been used by Karl Pearson in 1901 \cite{Pearson}. This analysis includes the decomposition of the eigenvalues of the covariance matrix.  \\
The average of the shapes is calculated as:

\begin{equation}
\overline{S} = \dfrac{1}{200} \sum _{i=1}^{200} S_{i} \ .
\end{equation}

By subtracting each sample shape from average shape matrix, the vertical vector $a_{i}$ can be calculated as: \\
 
\begin{equation}
a_{i}= vec(S_{i} - \overline{S}) \ .
\end{equation}

 These vertical vectors are used as the columns of matrix A, and the eigenvalue vectors of a covariance C is calculated by a singular value decomposition \cite{Press}. So we have:

\begin{equation}
A= (a_{1},a_{2}, \cdots , a_{200})=UWU^{T}
\end{equation}
and
\begin{equation}
C=\dfrac{1}{200}AA^{T} = \dfrac{1}{200}UW^{2}U^{T},
\end{equation}

$Vec(S)$ will change matrix $S$ to a column vector by concatenating its columns in a vertical order. 200 columns of orthogonal matrix $U$  are eigenvectors of the covariance matrix $C$ and $\sigma_{i}^{2}=\dfrac{\lambda_{i}^{2}}{200}$ is its eigenvalues which $\lambda_{i}$'s are diagonal elements of matrix $W$ that are arranged in decreasing order. The $i$th column of $U$ is shown by $U_{0,i}$ , and the principle component of $i$ will be changed to a $3\times n$ matrix by $ S^{i}=U_{0,i}^{(3)}$. The notation $a_{m \times 1}^{(n)}$, changes the $m \times 1$ vector a to a $n \times (m/n)$  matrix \cite{Minka}. \\

Now, instead of describing a new shape or texture as the linear combination of the samples as equation \ref{equation5}, they can be expressed by the linear combination of $n_{s}$ shapes and $n_{t}$ textures as principle components:

\begin{equation}
S=\overline{S} + \sum _{i=1}^{n_{s}} \alpha _{i} S^{i},
\end{equation}

and

\begin{equation}
T=\overline{T} + \sum _{i=1}^{n_{t}} \beta _{i} T^{i} \ .
\end{equation}

Therefore, the combination of an arbitrary number of shapes and textures for constructing 3D morphable faces is used. \\
The 3DMMs should be so limited that improbable faces could be rarely sampled. \\

Experiments on the real data shows that supposing Gaussian distribution on the face when sufficient information in the extraction is available, the result will be satisfying. Anyhow, supposing Gaussian distribution is unexpected, so Patel and Smith \cite{Patel} presented another assumption. They observed that the length of parameters' vectors has Chebyshev distribution. In other words, the real faces are on the spherical manifolds. In the model, the average model, which its vector length is zero, is improbable, and the faces should have a certain level of distinction. \\

\section{Extracting 3-D Shape and Texture From 2-D Face Images} \label{section:3}

\begin{figure}[!t] 
\centering
\includegraphics[width=3.5in]{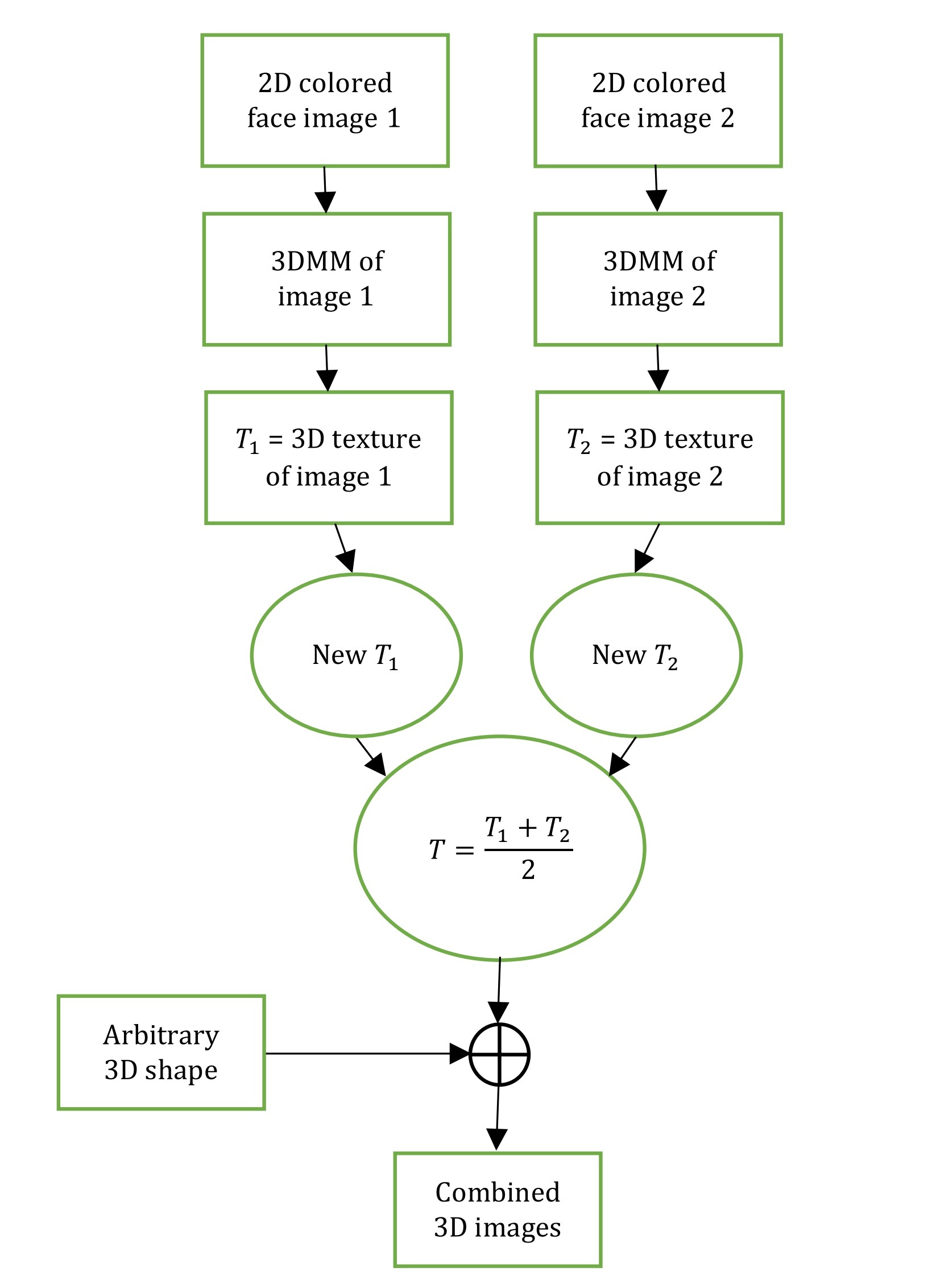}
\caption{Flowchart of extracting 3-D shapes and textures and combining them.  }
\label{photo:a}
\end{figure}

In this section extracting three-dimensional shapes and textures of 2-D colored images will be explained and we would show that how we can very well combine two of them.  The flowchart of this procedure is shown in Fig. \ref{photo:a}. 

\begin{figure}[!t] 
\centering
\includegraphics[width=3.5in]{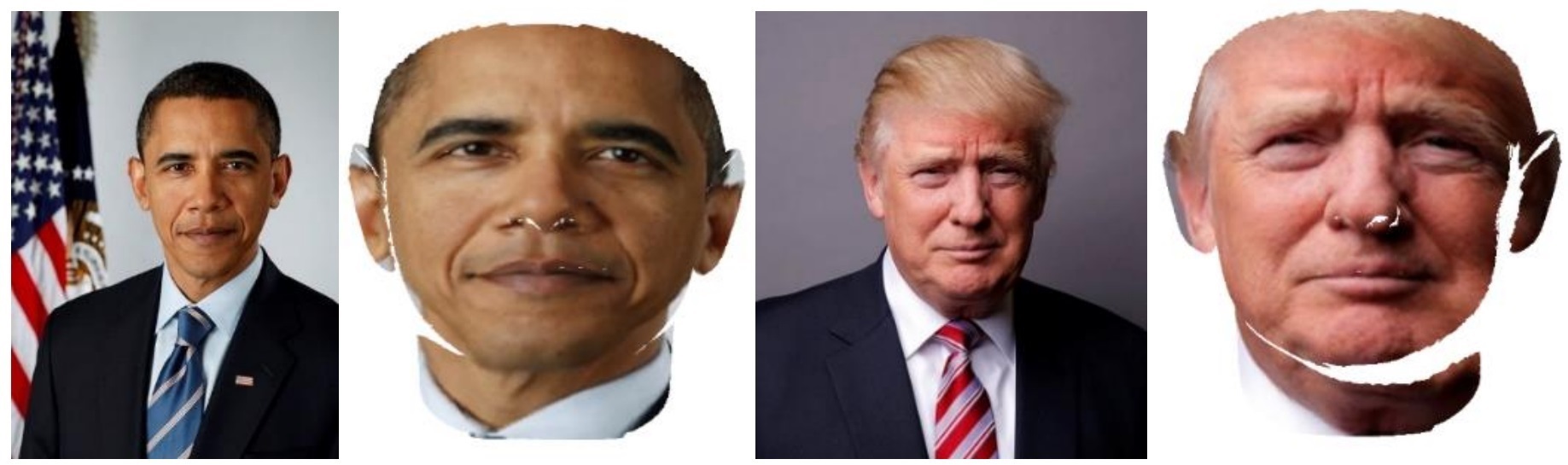}
\caption{2-D face images and their obtained 3-D models.}
\label{photo:b}
\end{figure}

Bas \cite{Bas} used also Basel database for fitting 2-D images on the 3-D model which fitting $N$ 2-D 
$X_{i}= \begin{pmatrix}
x_{i} \\
y_{i}
\end{pmatrix}  $
points for
$i=1,\cdots ,N$
 to the 3D 
$V= \begin{pmatrix}
 u \\
 v \\
 w
 \end{pmatrix}$
 vectors on the model for getting the best shape and pose which can minimize the Euclidean distance of the $N$ $X_{i}$ points from scaled orthogonal projection of four components $(V,R,t,s)$ which $R$ is a 3 by 3 rotation matrix of real numbers, $t$ is an translation 2-tuple of real numbers, and also $s$ scale is a real number which is a positive real number, are three pose parameters. By fitting 2-D image on a 3-D image, the 3-D shape $S$ and texture $T$ of the 3-D face are obtained as equation \ref{equation1} and \ref{equation2}
 which by using shapes and textures, the 3-D faces can be constructed. Now suppose that two 2-D face images, which can be chosen completely by accident, and their obtained 3-D images are as Fig. \ref{photo:b} .

%\begin{figure}[h!]
%\scalebox{.45}{\includegraphics{photo1.eps}}  
%\label{photo:11}
%\end{figure}

 The shape and texture of Obama's face are $S_{1}, T_{1}$, respectively, and the shape and texture of Trump's image are $S_{2} . T_{2}$ , respectively. \\

\begin{figure}[!t]
\centering
\includegraphics[width=1.7in]{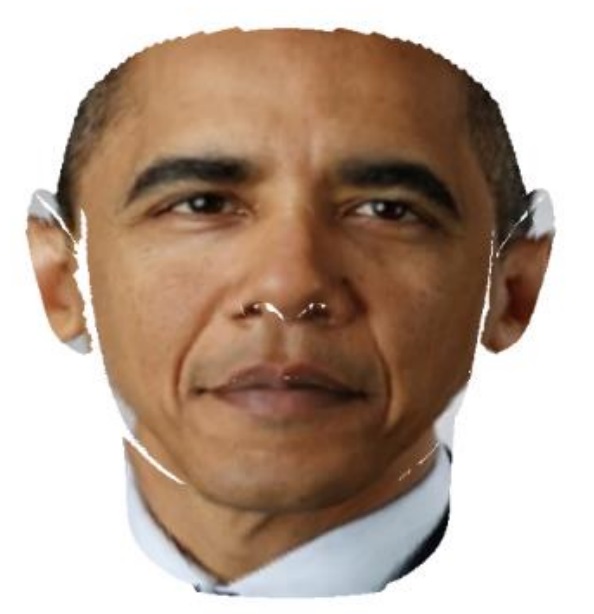}
\caption{Obama's texture on Trump's shape matrix}
\label{photo:c}
\end{figure}

Now, if we use Obama's texture which is a 3-D matrix $T_{1}$, and Trump's 3-D shape matrix which is shown by $S_{2}$, the Fig. \ref{photo:c}  will be obtained.\\

As it can be seen in the Fig. \ref{photo:c}, the 3-D shape for detecting a person's face is not so important, and only the face's texture for reconstructing human face is sufficient. For this reason, if we want to steganography a 3-D face image, it would be better to ignore half of the information, which is the shape matrix and only use the texture matrix which will be described in the section \ref{stegagnography}. \\

\begin{figure}[!t]
\centering
\includegraphics[width=3.3in]{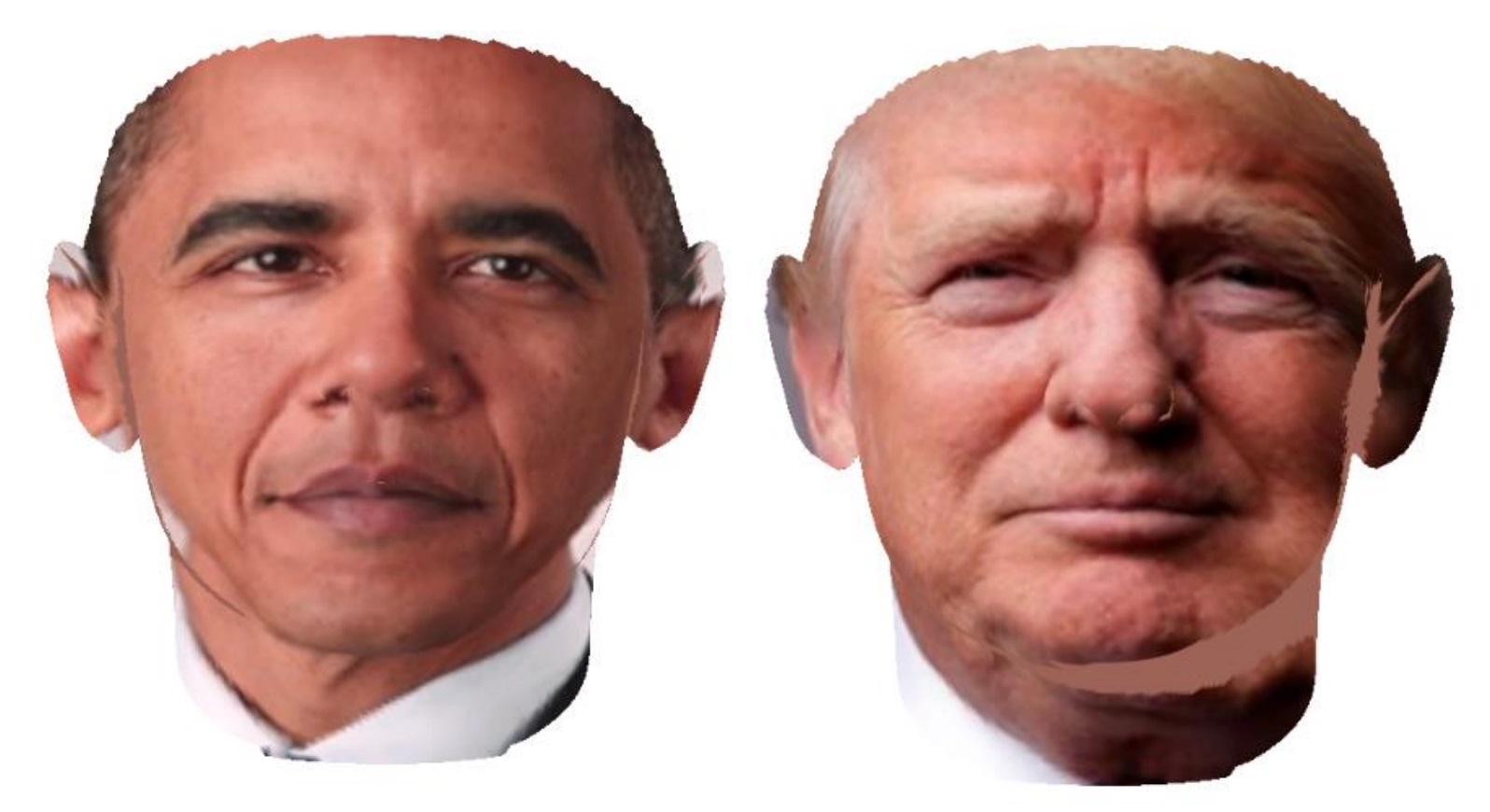}
\caption{Obama's and Trump's transformed shapes and textures as equations \ref{equation13}}
\label{photo:d}
\end{figure}

Moreover, if we want to combine the obtained 3-D images from the 2-D images with each other, an appropriate way is using coefficients of principle components. For $53490\times 3$ dimensional shape matrix $S_{1}$, by using $PCA$ with Alternating Least Squares $(ALS)$ algorithm \cite{Kiers} we have:

\begin{equation}
 [coeff S_{1}, scoreS_{1} ,\mu ]=pca (S_{1})
\end{equation}
that $3\times 3$ dimensional coefficient $coeffS_{1}$ is obtained which its columns are arranged in an increasing order and include any of the 3 principle component coefficients. The $53490\times 3$ dimensional score matrix is also shown by $scoreS_{1}$. The $1\times 3$ dimensional average of any variables' matrix is shown by $\mu_{S_{1}}$. In the same way, by principle component analysis and using $ALS$ algorithm, it can be done on the $S_{2}, T_{1} ,T_{2}$. Now, if the new shapes and textures are obtained as equations \ref{equation13}, the 3D face models of Fig. \ref{photo:d} can be constructed.

\begin{flalign} \label{equation13}
& T_{1} = scoreT_{1}\ast coeffT_{2}^{'} + repmat(\mu _{2},53490,1) , &&\\\nonumber
& T_{2} = scoreT_{2}\ast coeffT_{2}^{'} + repmat(\mu _{2},53490,1), &&\\\nonumber
& S_{1} = score S_{2} \ast coef S_{2}^{'} + repmat(\mu S_{2},53490,1), &&\\\nonumber
& S_{2} = score S_{2} \ast  coef S_{2}^{'} + repmat(\mu S_{2},53490,1) \ .  &&
\end{flalign}

\begin{figure}[!t]
\centering
\includegraphics[width=1.8in]{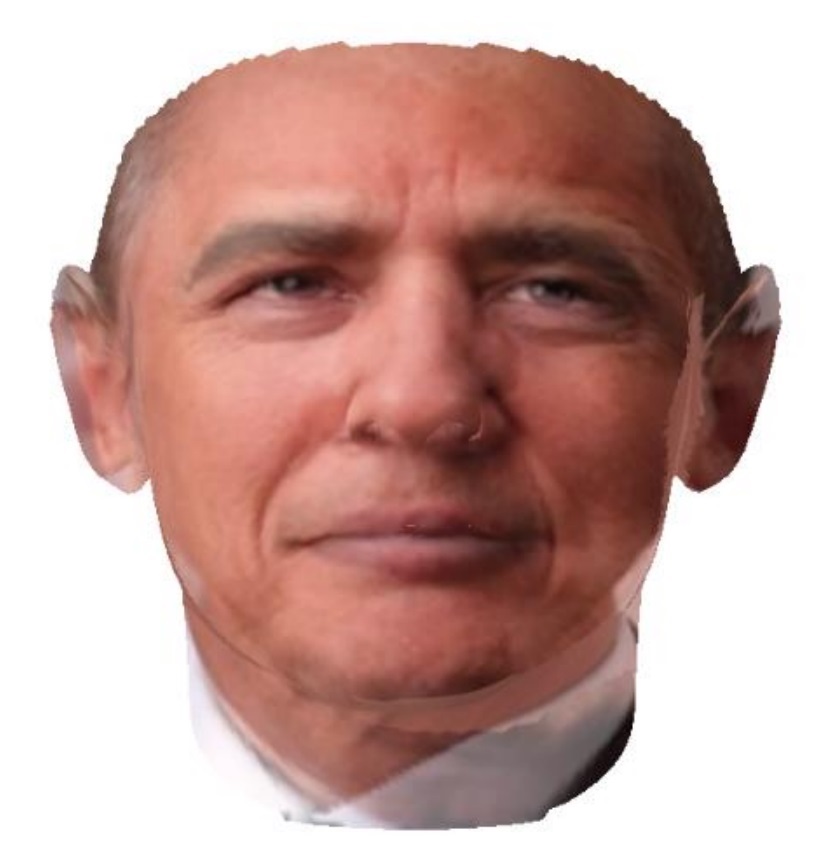}
\caption{Combined Obama and Trump 3-D models}
\label{photo:e}
\end{figure}

To construct the left face in Fig. \ref{photo:d}, the new shape and texture matrices $S_{1}, T_{1}$ have been used, and to construct the right face in Fig. \ref{photo:e}, the new shape and texture matrices $S_{2}, T_{2}$ have been used.  \\
Now by the transform which has been applied in \ref{equation13}, only by taking a simple average in \ref{equation14}, we can see a satisfactory behavior of two politicians as the face in the Fig. \ref{photo:e}.

\begin{equation} \label{equation14}
\begin{aligned}
S_{m}=(S_{1}+S_{2}) /2 \\
T_{m}=(T_{1}+T_{2}) /2
\end{aligned}
\end{equation}

\begin{figure}[!t]
\centering
\includegraphics[width=2.8in]{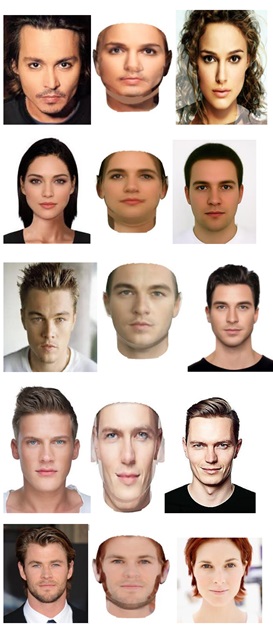}
\caption{3-D combining some arbitrary face images}
\label{photo:f}
\end{figure}

In Fig. \ref{photo:f} some other input 2-D colored face images and their combined 3-D models are shown.  

\section{ 3-D Face Image Steganography} \label{stegagnography}

As it is shown in the previous section, what is essential in the constructing of a 3-D face images and needs to be masked, is half of the face information videlicet, the texture of the 3-D face image, and by using any other shape the face can be reconstructed significantly same as the original one. \\

As it has been said, a texture matrix is a $53490\times 3$ dimensional matrix. To hide this 3-D image, we add some zeros to the end of this matrix to obtain a $53824\times 3$ dimension matrix which we have $232 \times 232 =53824$. By using the second dimension of the resulted matrix, we change it to three vertical $53824 \times 1$ vectors, and after that we reshape each of these vectors to a $232\times 232$  square matrix as the symbols of red, green, and blue colors, respectively. The values of these matrices which might not be number will be replaced by zero (also for better results, other techniques could be used by adjacent points to recover the missing texture points ). Now by using singular value decomposition, any of the colors is decomposed to a 3-tuples like $(U_{t} ,S_{t} ,V_{t})$, by subindices $i=r, g, b$ which are related to their colors. \\

Also, a colored image is used as the cover image which its size will be changed into to a $464\times 464\times 3$ dimensional matrix that we show it by $C$ ( the cover matrix can also be any other 3-D texture matrix that in this way we do as mentioned above). The matrix $C$, by using a single-level discrete 2-D wavelet transform, is decomposed to a 4-tuple $(CA,CH,CV,CD)$ by same $232\times 232\times 3$ dimensions which $CA$ is the approximation coefficient matrix which has low frequency, and 3 details coefficient matrices by the names horizontal matrix $CH$, vertical matrix $CV$  which has medium frequency, and diagonal matrix $CD$ which has high frequency. Now, because of the third dimension of the $CD$, we decompose it to 3 $CD_{r}, CD_{g}, CD_{b}$ matrices by $232\times 232$ dimension for the symbol of red, green, and blue colors, respectively. In the following, by using singular decomposition, each color matrices, like the matrices which were obtained by texture, is decomposed to 3-tuples $(Uc,Sc,Vc)$ by subindices belongs to the three colors. \\

Now, for red, green, and blue colors, the obtained singular value matrix from the cover image is replaced by the $S_{c_{i}}$ of that color $i$, added to the tenth of $S_{t_{i}}$, for example for the red color we can obtain new value of
\begin{equation} \label{equation15}
CD_{new_{r}}=Uc_{r} \times (Sc_{r}+0.1 \ast St_{r})\times Vc_{r}^{T}.
\end{equation}

\begin{figure}[!t]
\centering
\includegraphics[width=3in]{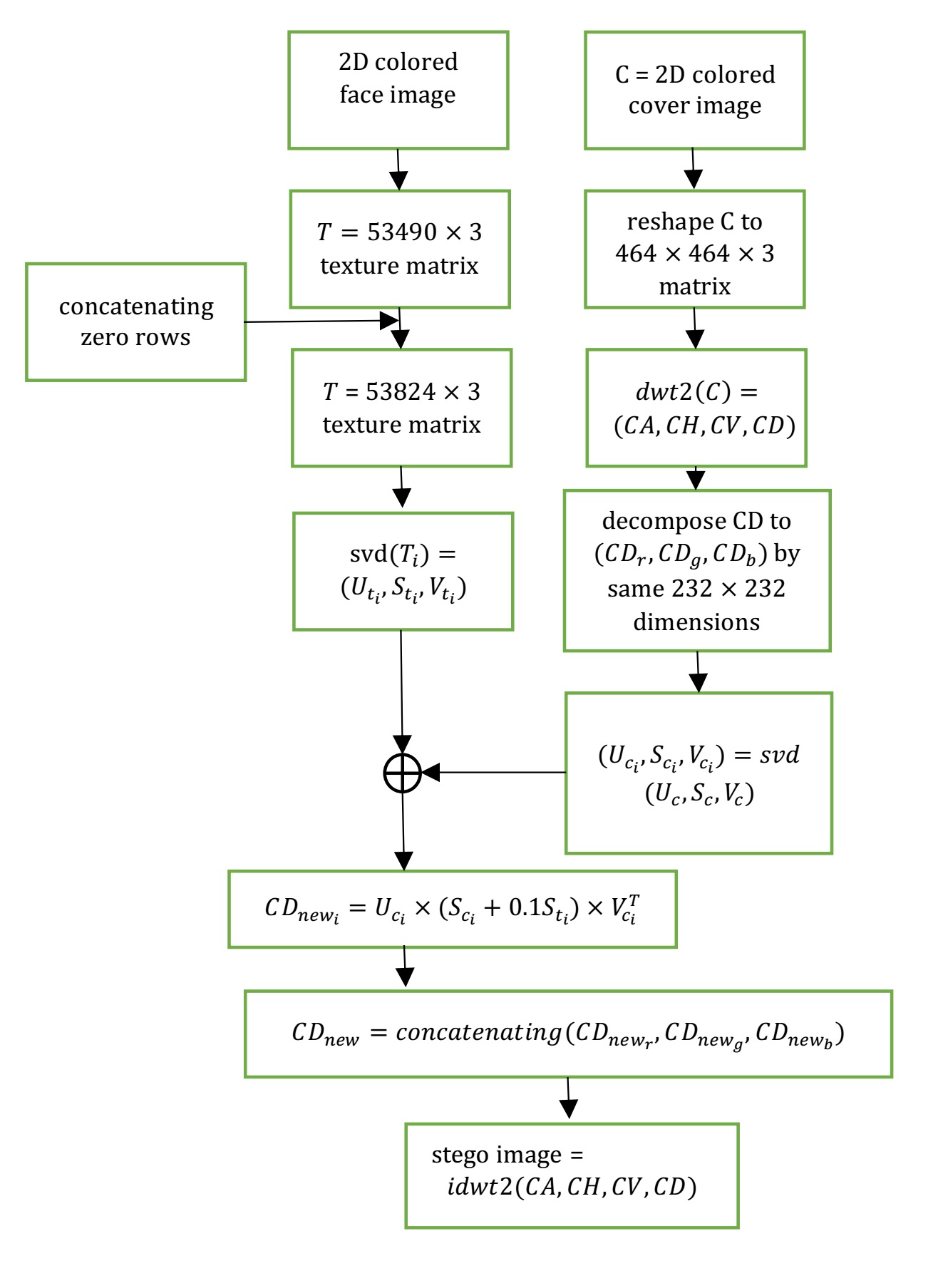}
\caption{Steganography of 3-D texture of face images which $i$ is equal to three red, green, and blue colors.}
\label{photo:g}
\end{figure}

Then by concatenation of these 3 red, green, and blue matrices, the new value of $D_{new}$ will be obtained and by inverse wavelet decomposition on the 4-tuple $(CA, CH, CV , CD_{new})$ the stego image will be constructed.  The procedure of steganography which is described in this section can be seen in Fig. \ref{photo:g}. \\

\begin{figure}[!t]
\centering
\includegraphics[width=3in]{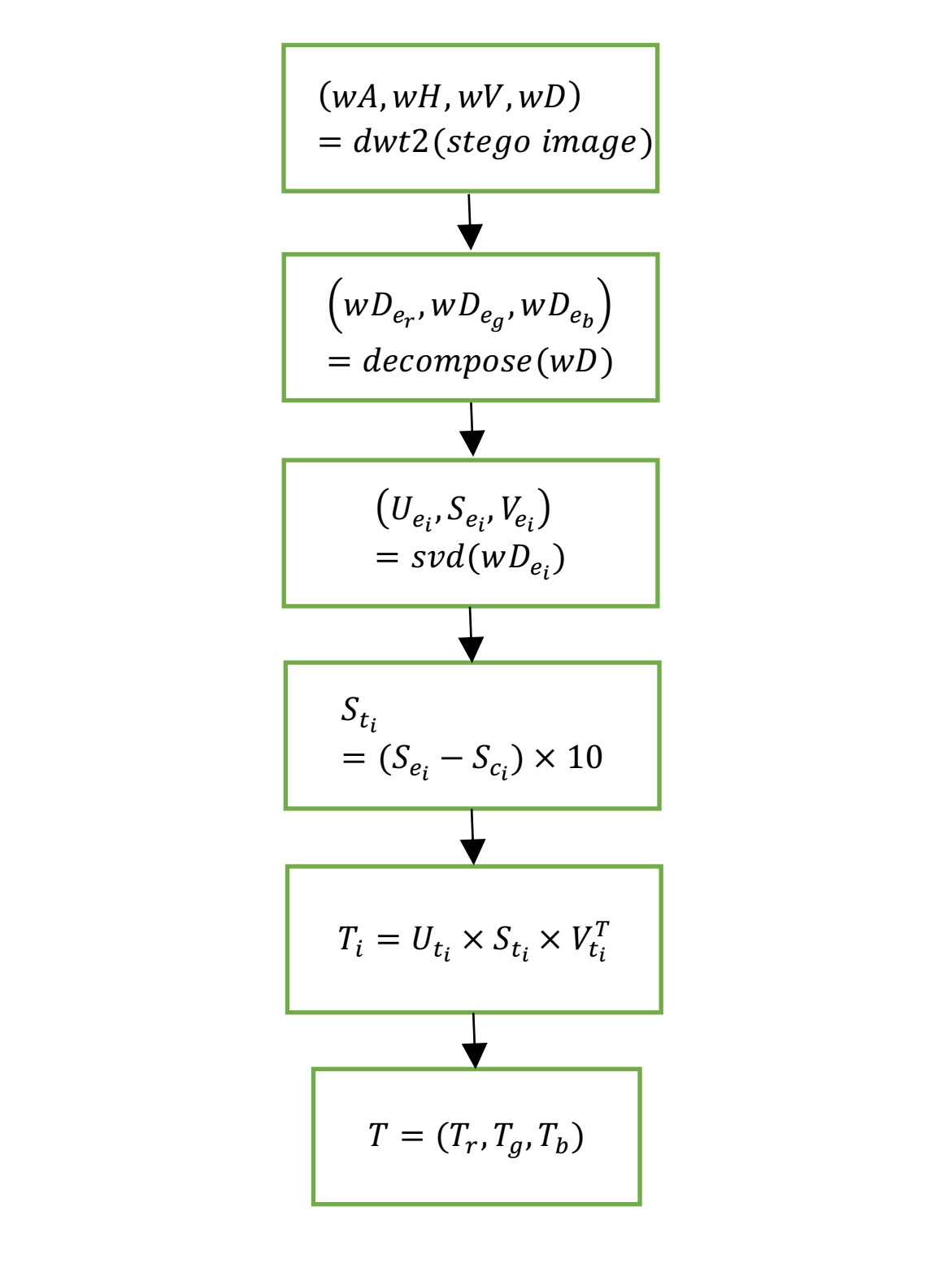}
\caption{Extracting hidden texture for constructing 3-D face image which $i$ is equal to three red, green, and blue colors.}
\label{photo:h}
\end{figure}

To extract the hidden image we should do as the inverse of abovementioned process, and at first 2-D wavelet decomposition for rebuilding 4-tuple $(wA,wH, wV, wD)$ is performed on the stego image. Because of the third dimension of $wD$, we decompose $wD$ to the three red, green, and blue color images, and on any color by singular value decomposition, the 3-tuple $(U_{e_{i}} ,S_{e_{i}},V_{e_{i}})$ by three subindices $i=r, g, b$ which belong to any color, is obtained which for calculating $S_{t_{i}}$ we should subtract that value of $S_{c_{i}}$ from obtained singular value $S_{e_{i}}$ and multiply it to 10 ( the inverse of what mentioned before). Next, the new texture column components from the result of $U_{e_{i}}\times S_{t_{i}}\times V_{e_{i}}^{T}$ for each color will be calculated. This procedure is shown in the Fig. \ref{photo:h}.  \\

\begin{figure}[!t]
\centering
\includegraphics[width=3.5in]{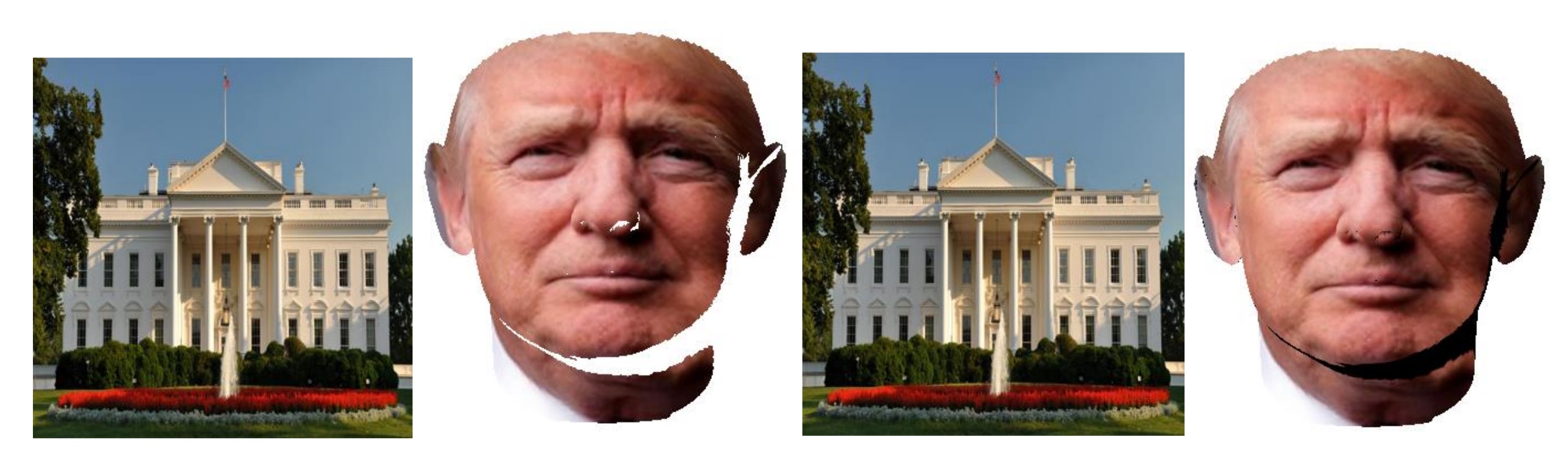}
\caption{Steganography of 3-D texture which from left to right are the original cover image, 3-D used original image, stego image, and desired extracted the image, respectively.}
\label{photo:i}
\end{figure}

We can implement the resulted texture on any shape to reconstruct the desired 3-D face image. As an example, the images for the mentioned process are as Fig. \ref{photo:i} which from left to right are the original cover image, 3-D used original image, stego image, and desired extracted the image, respectively. A known used criteria for measuring image quality is the peak of the signal to noise ratio \cite{Rafael}. The peak of the signal to noise ratio of two original cover image and stego image of Fig. \ref{photo:i} is $81.51$ (dB) by equation \ref{equation16} which MSE is the Mean Square Error and S is the maximum pixel value. A known used criteria for measuring image quality is peak of signal to noise ratio \cite{Rafael}. The peak of signal to noise ratio of two original cover image and stego image of Fig. \ref{photo:i} is $81.51$ (dB) by equation \ref{equation16} which MSE is the Mean Square Error and $S$ is the maximum pixel value. 

\begin{equation} \label{equation16}
PSNR=-10 log_{10}\dfrac{MSE}{S^{2}}
\end{equation}

\begin{figure}[!t]
\centering
\includegraphics[width=3.5in]{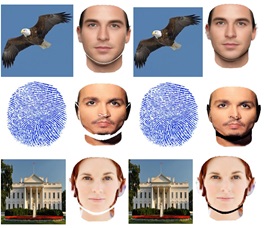}
\caption{Steganography and extracting of three other textures with PSNR equal to 79.4391, 81.3087, and 80.9286 dB, respectively. }
\label{photo:j}
\end{figure}

Also, some other covers and hidden images with their PSNR values are shown in the Fig. \ref{photo:j}. 
In all results, the destruction of cover image by hiding texture in it from our method is very insignificant which all the PSNR values are near to 80. A similar result is achieved by Chao \cite{Chao}.

\section{Conclusion}
Hiding the secretive information from the enemies' view in a way that they do not be at all skeptical to the existence of hidden messages helps to the safe communication. In this paper, by using the idea of extracting 3D face texture, this matrix is embedded in other images, and it has been indicated that how by using only half of the face information, which is texture, we can appropriately reconstruct face images. In addition, it has been shown that how by combining two textures and without any stricture on shape matrix, we can combine two faces pleasantly.

\end{document}